\documentclass[aps,prb,twocolumn,showpacs,groupedaddress]{revtex4}
\usepackage[latin9]{inputenc}
\usepackage{textcomp}
\usepackage{hyperref}
\usepackage{mathtools}
\usepackage{amsmath}
\usepackage{bbm}
\usepackage[usenames, dvipsnames]{color}
\makeatletter
\renewcommand*\env@matrix[1][\arraystretch]{%
  \edef\arraystretch{#1}%
  \hskip -\arraycolsep
  \let\@ifnextchar\new@ifnextchar
  \array{*\c@MaxMatrixCols c}}
\makeatother
\usepackage{amssymb}
\usepackage{graphicx}
\usepackage{setspace}
\usepackage{esint}
\usepackage{url}
\usepackage{appendix}
\usepackage{indentfirst}
\usepackage{braket}
\usepackage{multirow}
\usepackage{float}
\usepackage[font={footnotesize}]{caption}
\usepackage{pdfpages}
\usepackage[normalem]{ulem}
\DeclarePairedDelimiter\abs{\lvert}{\rvert}%
\DeclarePairedDelimiter\norm{\lVert}{\rVert}%
\makeatletter
\let\oldabs\abs
\def\abs{\@ifstar{\oldabs}{\oldabs*}}
\let\oldnorm\norm
\def\norm{\@ifstar{\oldnorm}{\oldnorm*}}
\makeatother

\definecolor{mypink}{RGB}{219, 48, 122}

\begin{document}
\title{Topological states at the (001) surface of SrTiO$_\text{3}$}
\author{Manali Vivek, Mark O. Goerbig and Marc Gabay }

\affiliation{ Laboratoire de Physique des Solides, Universit\'e Paris-Sud 11, Universit\'e Paris Saclay, CNRS UMR 8502, 91405 Orsay Cedex, France }
\date{\today} 

\begin{abstract}
Defect-free SrTiO$_3$ (STO) is a band insulator but Angle Resolved Photoemission Spectroscopy (ARPES) experiments have demonstrated the existence of a nanometer thin two-dimensional electron liquid (2DEG) at the (001) oriented surface of this compound. The bulk is a trivial insulator, but our theoretical study reveals that the parity of electronic wavefunctions in this 2DEG is inverted in the vicinity of special points in reciprocal space where the low-energy dispersion consists of four gapped Dirac cones with a tilted and anisotropic shape. This gives rise to linearly dispersing topological edge states at the one-dimensional boundary. We propose to probe these modes by measuring the Josephson radiation from gapless bound Andreev states in STO based junctions, as it is predicted that they display distinctive signatures of topology.
\end{abstract}

\maketitle
\section{Introduction}
Material science has sparked a recent interest in the research community with the emergence of certain innovative compounds with desirable transport properties. Of equal interest are the compounds that are endowed with interfacial or surface states whose properties vastly differ from their bulk parent compounds. For example, one finds metallicity confined at the boundary between  two wide band gap bulk insulators, LaAlO$_3$ (LAO) and SrTiO$_3$ (STO) \cite{ohtomo2004high}. Such unusual behavior "at the edge" is truly an interfacial effect with the bulk of the two materials playing no role. This might be reminiscent of topological matter where unconventional states develop at the edge \cite{Ren2016reviewisotop}.

Apart from these unexpected interfacial effects, ARPES studies have revealed that even the surface of (001) oriented STO, in the presence of oxygen vacancies, harbours a metallic state while the bulk remains insulating. It is to be noted that bulk STO is a band insulator in the absence of these vacancies. The metallic state or the 2D electron gas (2DEG) at the surface has a band structure which is very different from its 3D counterpart. It is not simply a 2D cut of the bulk band structure, as its shows essentially no dispersion in the direction perpendicular to the 2DEG \cite{plumb2014mixed}.

In addition to the metallic state present at the surface, theoretical proposals advocating the presence of topological states in (111) oriented transition metal oxides have also been made for heterostructures \cite{Okamoto2011topol,liang2013electrically,ruegg2011,ruegg2013} and subsequently surfaces, where a metallic state already exists \cite{bareille2014,rodel2014orientational} . According to ARPES measurements, (111) oriented STO and KTaO$_3$ (KTO) have a stongly confined 2DEG. The geometry of the conducting paths of the carriers in the 2DEG for this particular orientation is a honeycomb lattice.  If its extent is limited to two unit cells, a theoretical analysis of the parity of eigenstates suggests a Z$_2$ topological state provided that $E_f$ is close enough to the position of the topological band gap predicted. However, these two requirements, of limiting the gas to two unit cells and $E_f$ being close to the topological gap, are not met experimentally.

In this paper, we uncover a topological character of the conducting electronic states of the 2DEG present at the (001) oriented surface of STO. These features are specific to the surface and have no bearing on the bulk properties of the system. Accordingly, edge states are expected at the 1D boundary of this 2D layer. 

While some of the basic ingredients known to be relevant to the physics of topological insulators are also seen here, certain specific features in this system need to be underlined, namely the occurence of several bands of different orbital character and the impact of the confining potential on the orbital ordering.

The scope of the present paper is to bring to light the existence of the 1D topological edge states, stemming from a non trivial Z$_2$ number for the eigenstates in the Brillouin Zone (BZ) in the 2DEG observed in ARPES at the (001) surface of STO.

The first section of our paper summarizes the main experimental findings pertinent to the electronic structure of the 2DEG at the surface of STO which show a dominant $d$ character of the conduction bands, hint at the two dimensionality of the gas, show signatures of the confinement potential and the electric field associated to the Rashba spin-orbit coupling present at the surface. In the second section, we propose a realistic theoretical model to incorporate all the elements listed above, in an energy range of 300 meV around $E_f$, which causes us to neglect the higher Sr states and the $e_g$ states of Ti. After, we explicit each term which enters the model, confinement, bulk spin-orbit coupling and its combination with orbital mixing at the surface, giving rise to an effective Rashba coupling.  We then ascertain the accuracy of the model by overlaying the theoretically determined energy and momentum dispersion curves onto the experimental ARPES data. Near $E_f$, the energy bands have a tilted, Dirac-like 
geometry in the vicinity of four symmetry-related points in the Brillouin Zone (BZ). We argue that these points could be viewed as effective $\Gamma $ points where again the orbitals are degenerate, as is observed in the cubic bulk structure without Spin-Orbit Coupling (SOC).
In the next section, we present the eigenstates in the vicinity of the DP and we discuss the symmetries they posses and the winding of the phase of the wave-functions on constant energy contours. Branch cuts of the phases connecting pairs of time-reversed Dirac partners are also shown as well as the parity inversion occuring in the gapped region.
An edge state analysis is provided in section four and we find linearly dispersing 1D edge states at the boundary of the 2D surface, by mapping our model to a simpler Bernevig-Hughes-Zhang (BHZ) model \cite{bernevig}.
The last section includes the conclusions and perspectives where we discuss the potential impact of the magnetic states in the 2DEG. We also discuss possible experiments to provide a fingerprint of these 1D topological states. It is noteworthy to mention that since these states lie a few meV below the $E_f$, it is easier to access these states as compared to other transition metal oxides, notably (111) oriented KTO and STO.

\section{Experimental background}
STO is known experimentally to be a trivial band insulator in the bulk, in the absence of defects. In the bulk for the cubic phase, the conduction $t_{2g}$ states $d_{xy},d_{yz},d_{xz}$ are degenerate at the $\Gamma$ point and are unoccupied, giving rise to its insulating behaviour. After the addition of bulk SOC, the $t_{2g}$ states are spin-orbit coupled and the $\Gamma_7^{+}$ spin split state is raised to about 25-30 meV above the degenerate $\Gamma_8^{+}$ states at the $\Gamma$ point. Below 100 K, there also exists a tetragonal distortion which might remove the degeneracy of the bands but it is largely dominated by SOC. However, away from $\Gamma$, the bands essentially recover their pure orbital character, irrespective of the perturbation. In 2011, interest was revived in STO, when a 2DEG was found to exist at its (001) surface \cite{santander2011two,Baumberger2011}. Its origin is ascribed to oxygen vacancies confined to the boundary. To characterise it theoretically, the electronic structure, derived 
from spin-integrated ARPES experiments, was interpreted in terms of sub-bands.  The sub-band structure was thought to be caused by the confinement potential, modelled by a triangular quantum well, which pins the 2DEG at the surface. This confinement potential also has another important effect; it overcomes the bulk SOC effect to lower the $d_{xy}$ states by about 200 meV below $E_f$, giving metallicity and raising the $d_{xz},d_{yz}$ states above the $d_{xy}$, thus, reversing the order of the bands. The two dimensionality of the gas was also revealed by the lack of dispersion in the $\hat{z}$ direction. The dominant orbitals in the conduction bands were shown to be the $t_{2g}$ $d$ states of Ti and those in the valence bands were shown to be O $2p$ states.

Subsequently, Spin-Polarised ARPES \cite{santander2014} was performed and spin polarisation was found for the two sub-bands seen earlier in spin-integrated ARPES. The data was interpreted as spin split bands with oppositely winding spin chiralities at the Fermi surface. The spin textures observed were a consequence of the surface Rashba effect.

An issue raised with the detection of spin polarised bands was the presence of internal magnetism and spin-split bands \cite{santander2014,Baumberger2016nomagnetism}, which is one we will discuss further when we probe our system for possible topological states.
In the next section, we will show the modelling of the bands.

\section{Modeling the 2DEG}

 In order to obtain an appropriate low-energy model for the electronic states of the 2DEG, we follow a two-step procedure. First, we construct a tight-binding Hamiltonian that includes the relevant terms capturing the physical properties of the material. In addition to an effective hopping between Ti atoms for each of the three $t_{2g}$ $d$ orbitals, we take into account the confinement close to the surface of STO and in a direction perpendicular to it, denoted by $z$. Confinement quantizes the dynamics along $z$. Furthermore, we include the SOC as well as orbital mixing contributions. Second, once the structure of the model is set by these physical considerations, we determine the values of the effective tight-binding parameters from a fitting to available ARPES spectra.
In the insulating bulk, the conduction band manifold has a dominant $d$ orbital character of the charges which leads to an effective Ti $t_{2g}$ and O $2p$ Hamiltonian where hopping is between neighboring Ti and O sites. The directional anisotropy of the bondings implies that the hopping amplitude of $d_{yz}$ carriers is small in the $\hat{x}$ direction, denoted by \textit{t$_{\text{h}}$}, but large in the $\hat{y}\text{ and }\hat{z}$ directions denoted by \textit{t} and is degenerate at the $\Gamma$ point, as can be seen in Fig. 1b) of \onlinecite{santander2011two}. In the presence of bulk SOC, the degeneracy of the bands at $\Gamma$ is lifted. To construct a low-energy Hamiltonian, one needs to take into consideration the orbitals which can play an important role at the surface near $E_f$. The reduction of the Hilbert space from a hybridization between O $2p$ states and Ti $t_{2g}$ states is a viable choice as the difference in energy between the $e_g$ and the $t_{2g}$ manifolds is of the order of 2.5 
eV. The other states, for example, the core states of Ti, O and Sr lie far below the $E_f$ and other unoccupied bands of Ti, O and Sr lie much higher up in energy. Thus, a low energy Hamiltonian can be constructed from the reduction of the Hilbert space from a hybridization between O $2p$ states and Ti $t_{2g}$ states and then to an effective model with hopping from Ti to Ti sites. This bulk modelling can be used as a support when we move to the surface but it is important to note that at the surface, new terms in the Hamiltonian will completely change the allure of the bands. This modelling of the conduction bands is also in line with the dominant $d$ orbital charater of the conduction band as revealed in ARPES \cite{santander2011two,Baumberger2011} and confirmed in DFT \cite{Shen2012}. Mapping a tight-binding Hamiltonian onto the experimental and simulated data is a standard procedure used by the ARPES and DFT communities to extract the relevant energies and effective masses. In order to quantify the 
interplay between the orbital and spin degrees of freedom, we describe the ARPES determined bands with the help of simple tight-binding surface models, inclusive of kinetic, bulk spin-orbit and surface orbital mixing terms.  

 The directions chosen for the surface are $\hat{x}$ and $\hat{y}$ while the perpendicular to the surface is in the $\hat{z}$ direction.
The surface Hamiltonian $H_{\text{surf}}$ has the form 
\begin{equation}\label{ham}
H_{\text{surf/2DEG}}=H_{\text{kin}}+H_{\text{so}}+H_{\text{om}}.
\end{equation}
$H_{\text{kin}}$ is composed firstly of tight-binding bands derived from the effective Ti-Ti hopping as in the bulk. Secondly, the motion of the carriers within the 2DEG is constrained along $\hat{z}$ by the confinement potential, caused by the oxygen vacancies. 
 Confinement then yields two terms in the Hamiltonian -- the first, a global energy offset -$V_0$, \cite{santander2011two} affects all three $t_{2g}$ bands equally and the second, $\epsilon_{1/2}$, represents the two relevant sub-bands due to the $k_z$-quantization. 

At $\Gamma$, the sub-band value $\epsilon_2$ for $d_{xz}$ and $d_{yz}$ states is greater than the energy $\epsilon_1$ for the $d_{xy}$ state, reflecting the hierachy of hopping amplitudes along $\hat{z}$ in the bulk. The energies are thus, 
\begin{eqnarray}
\epsilon_{d_{yz}}=&2t(1-\cos k_y)+2t_{\text{h}}(1-\cos k_x)+\epsilon_2-V_0 \\
\epsilon_{d_{xy}}=&2t(2-\cos k_x-\cos k_y)+\epsilon_1-V_0 \\
\epsilon_{d_{xz}}=&2t(1-\cos k_x)+2t_{\text{h}}(1-\cos k_y)+\epsilon_2-V_0 
\end{eqnarray}

where the in-plane wave-vector components $k_x$ and $k_y$ are dimensionless (momenta times $a$, the lattice parameter). The $d_{xy}$ states are lowest in energy at $\Gamma$ and the orbital order is reversed as compared to that in the bulk. Consequently one observes several crossing points between one light ($d_{xy}$) and one heavy ($d_{xz}$ or $d_{yz}$) band in the BZ along $\Gamma$X and $\Gamma$Y. However, along $\Gamma$M and $\Gamma\bar{\text{M}}$, one gets special crossing points, where all three  orbital energy dispersions are equal denoted by $\Gamma_\text{dp}$. The addition of confinement breaks the degeneracy of the orbital energies at $\Gamma$ that one observes in the bulk cubic phase, but in a certain sense this degeneracy is restored at $\Gamma_{\text{dp}}$ bringing a new symmetry at the surface in (and four new "$\Gamma$" - like points). At $\Gamma_{\text{dp}}$, $\abs{k_x}= \abs{k_y}=k_c =2\arcsin{\left\{\sqrt{\left[\left(\epsilon_2-\epsilon_1\right)/\left(t-t_h\right)\right]}\right\}}$.

\begin{figure}[h]
  \includegraphics[width=0.5\textwidth]{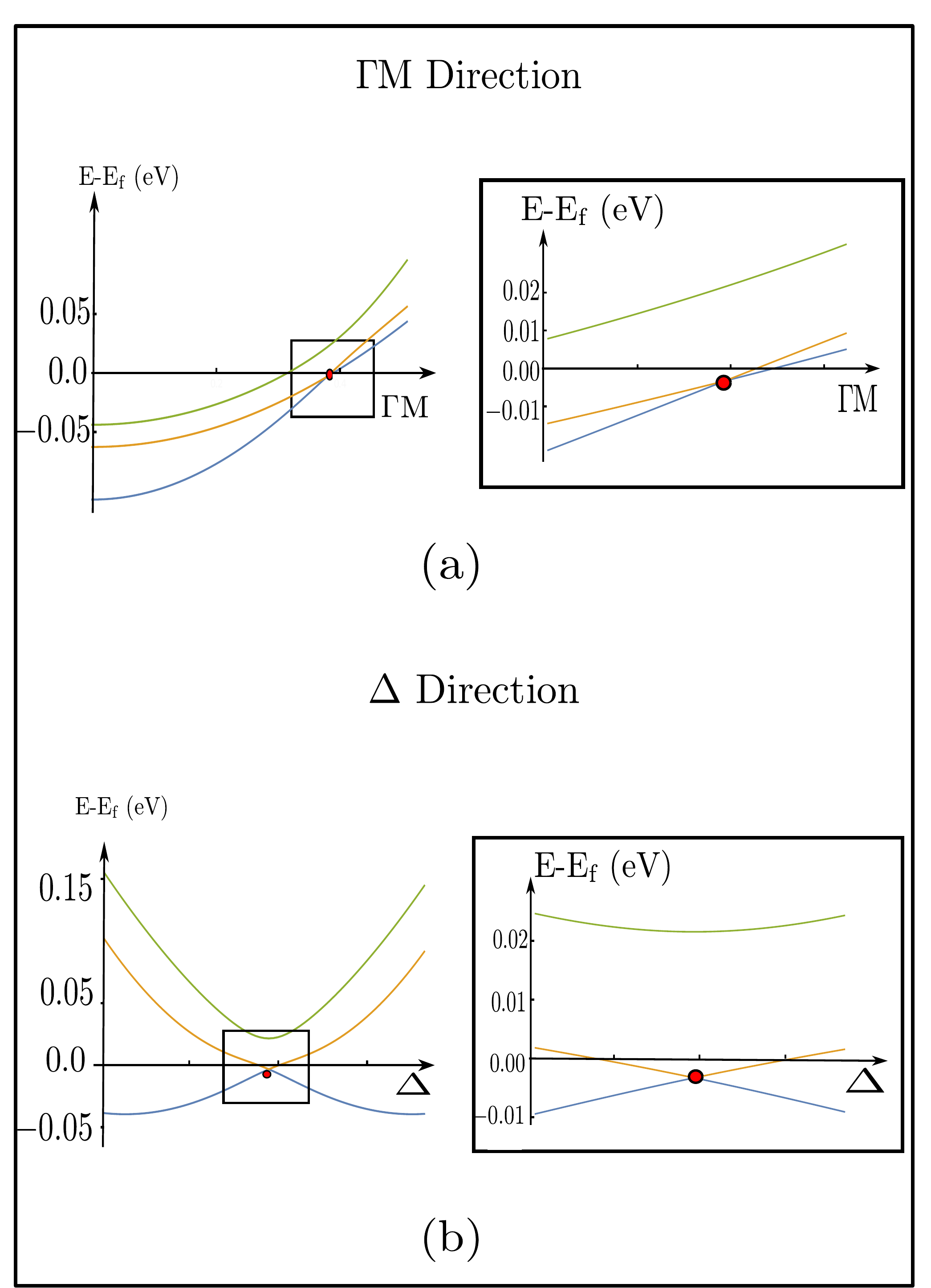}
  \caption{Color online. Spin orbital band structure in the first quadrant of the BZ, along $\Gamma$M and $\Delta$, (the direction perpendicular to  $\Gamma\text{M}$ and passing through  $\Gamma_{\text{dp}}$)  (a)-(b) without the orbital mixing term ($g=0$). A zoom of the boxed areas close to $\Gamma_{\text{dp}}$ (red dot) is shown to the right. }
  \label{fig1}
\end{figure}

The next contribution in Hamiltonian (\ref{ham}) is simply the bulk SOC due to the Ti atoms which entangles the spin and orbital degrees of freedom, the expression of which is given by
\begin{equation}
H_{\text{so}}=\lambda { \text{ }\bf L}\cdot{\bf S}
\end{equation}
At the $\Gamma$ point, confinement has already lifted the degeneracy of the bands and the bulk SOC does not produce a significant effect. However it is an  essential ingredient at the $\Gamma_\text{dp}$ points and results in making the band crossings partially avoided; one of the bands, $\Gamma_7^{+}$ becomes spin-orbit split off and which is raised by 25-30 meV above the other two degenerate $\Gamma_8^{+}$. The points at which the two degenerate bands still cross are the called Dirac Points (DP) situated at $\Gamma_\text{dp}$. They are four in total, one in each quadrant of the BZ and are related by C$_4$ and time reversal symmetry (see Fig. \ref{fig1}). In the vicinity of the DP, the orbital dispersion resembles Dirac cones.  Each of the Dirac cones is tilted and the tilt angle changes as one moves around $\Gamma_{\text{dp}}$. In the first quadrant, it is maximum along $\Gamma$M  but goes to zero in the direction denoted by $\Delta$, perpendicular to $\Gamma{\text{M}}$ and passing through $\Gamma_{\text{dp}
}$. The bulk SOC does not change the position of the crossing point and the Dirac cones are still found at $\Gamma_\text{dp}$. Because of time reversal symmetry, each spin orbital  band is twofold degenerate in energy. 
We note that, while the points $\Gamma_\text{dp}$ are situated in directions of high symmetry, their crystal symmetry is lower than that of the special points $\Gamma$ or $M$ and $\bar{M}$. This feature allows for the existence of a tilt of the Dirac cones which would have been otherwise excluded \cite{goerbigPRB,goerbigEPL}.

\begin{figure}[h]
  \includegraphics[width=0.5\textwidth]{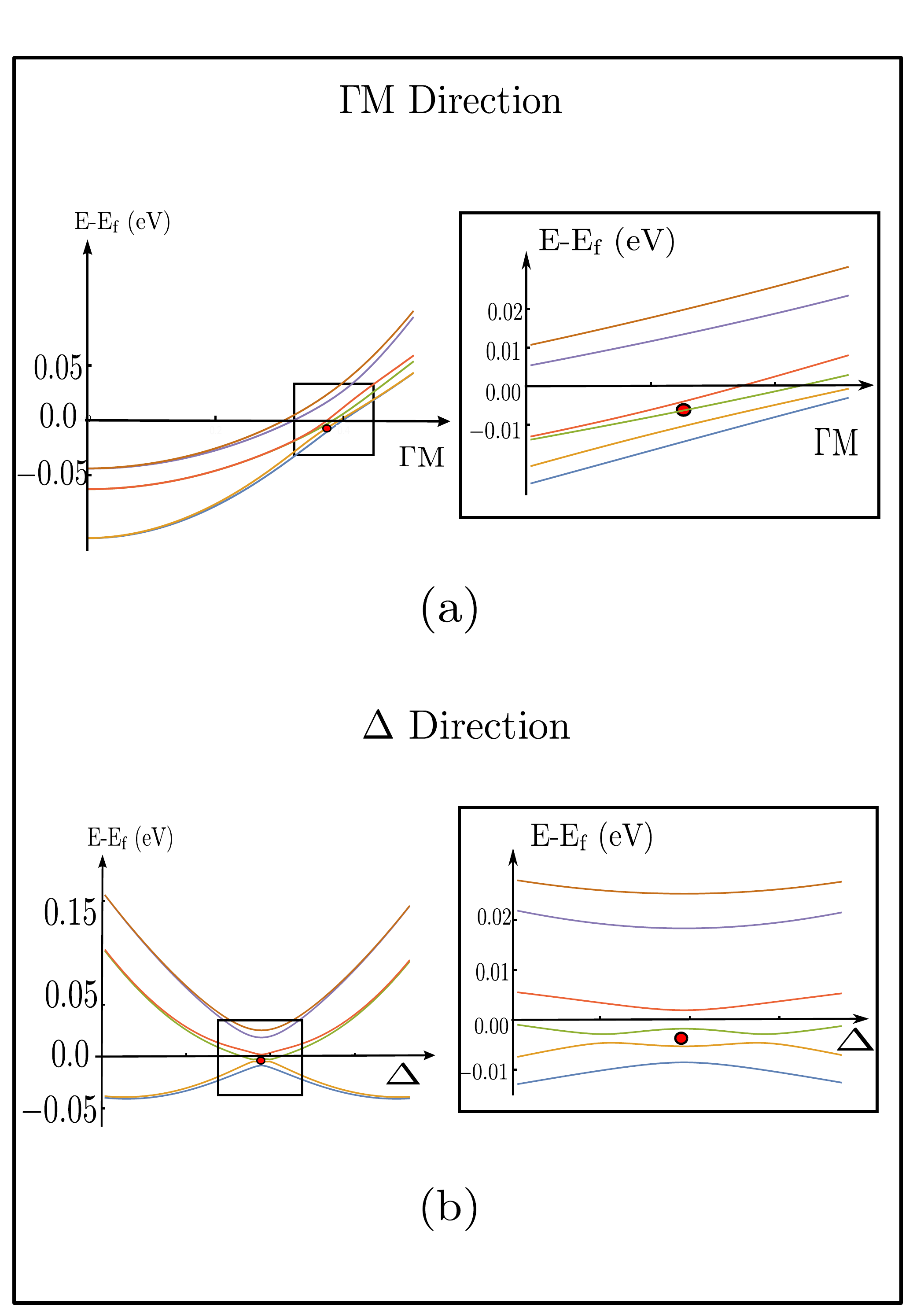}
  \caption{Color online. Spin orbital band structure in the first quadrant of the BZ, along $\Gamma$M and $\Delta$, (the direction perpendicular to  $\Gamma\text{M}$ and passing through  $\Gamma_{\text{dp}}$)   (a)-(b) with the orbital mixing term. Zoom of the boxed areas close to $\Gamma_{\text{dp}}$ (red dot) on the right. A band inversion is seen along $\Delta$ (b) but not along $\Gamma$M (a).}
  \label{fig2}
\end{figure}

The last term in Hamiltonian (\ref{ham}) is the orbital mixing (OM) term, which originates from the electrostatic and the bond angle perturbations that occur near the surface. Given the value of the carrier density of the 2DEG, and considering that the excess electrons in the gas originate from the vacancies, we may consider that the vacancies are fairly homogenously distributed across the surface. The OM term gives rise to a Rashba term whose  expression can be found in  [\onlinecite{khalsa2013theory,Zhong2013}]
\begin{eqnarray}
\Braket{xy;\sigma|H_{om}|yz;\sigma}=2 i g \sin k_x \\
\Braket{xy;\sigma|H_{om}|xz;\sigma}=2 i g \sin k_y
\end{eqnarray} where $g$ is the strength of the OM and $\sigma = \uparrow,\downarrow$ . This produces new coupling terms among orbitals which were earlier forbidden in the bulk by symmetry.
The experimentally relevant value of $g$ is at $E_f$. It is particularly large
whenever $E_f$ lies close to the gaps of the band structure (i.e. near the avoided band crossings). A combination of tight-binding and DFT calculations estimate the size of the Rashba-like term to be 5-10 meV which is consistent also with the experimentally determined value. The OM term mixes the spin-orbital states derived from $H_\text{kin}$ and $H_\text{SO}$ and lifts their degeneracies. In particular gaps develop in the spectrum at $\Gamma_{\text{dp}}$ as shown in Fig \ref{fig2}.  
In the basis $\Ket{d_{\beta};\sigma}$ ($\beta=(xy,yz,xz), \; \sigma = \uparrow,\downarrow$), with $\lambda^{'}=\lambda/3$,  $H_{\text{surf/2DEG}}$ reads
\begin{equation}\label{bigass_ham}
\left[
\begin{smallmatrix} 
\epsilon_{d_{xy}} & 2ig\sin k_x &2ig\sin k_y & 0& \lambda^{'}&-i \lambda^{'} \\
-2ig\sin k_x & \epsilon_{d_{yz}} & i\lambda^{'} & -\lambda^{'} & 0 &0   \\
-2ig\sin k_y & -i\lambda^{'}& \epsilon_{d_{xz}} & i\lambda^{'} & 0 &0 \\
0& -\lambda^{'}&-i\lambda^{'} & \epsilon_{d_{xy}}& 2ig\sin k_x & 2i g\sin k_y \\
\lambda^{'} & 0 & 0  & -2ig\sin k_x&\epsilon_{d_{yz}}& -i\lambda^{'}\\
i\lambda^{'} & 0 & 0 & -2ig\sin k_y & i\lambda^{'} &\epsilon_{d_{xz}}
\end{smallmatrix}
\right]
\end{equation}
It is to be noted that this Hamiltonian is for the same value of $k$ and its time-reversed block, degenerate in energy, also exists but will not be represented here for the sake of simplicity.
Excellent fitting of the ARPES energy dispersion curves is obtained with $a= 3.90$ \AA, $t=0.388 $ eV, $t_h=0.031$ eV, $\lambda =0.025$ eV, $g=0.02$ eV and $\epsilon_1-V_0=-0.205$ eV (1st $d_{xy}$ band) and $g=0.005$ eV and $\epsilon_1-V_0=-0.105$ eV (2nd $d_{xy}$ band), $\epsilon_2-V_0=-0.0544 $ eV for $d_{yz}/d_{xz}$ bands. With these parameters and no further adjustment, we overlay the tight binding bands onto the momentum dispersion curves at $E_f$ and obtain a very good agreement. DFT calculations on slabs with vacancies confirm the values of the bulk SOC and the Rashba SOC. 
Within the resolution of ARPES, one sees these crossing points \cite{plumb2014mixed, Roedel2016capping}. DFT calculations for slabs with vacancies also produce dispersion suggestive of the 3 band crossing \cite{Shen2012,Jeschke2015}. A recent experimental report of the bands in LAO-STO \cite{Strocov2016}  also shows  two types of crossings but it is to be noted that the resolution of the ARPES measurement in this heterostructure is less than that achieved at the STO (001) surface. Gating of the samples is also a simple way to tune the energies of the surface states in order to exhibit $\Gamma_{\text{dp}}$ -type crossings. In the next section, we will detail the impact of all these terms on the surface spectrum and the surface eigenstates.
 
\begin{figure}[h]
\includegraphics[width=0.5\textwidth]{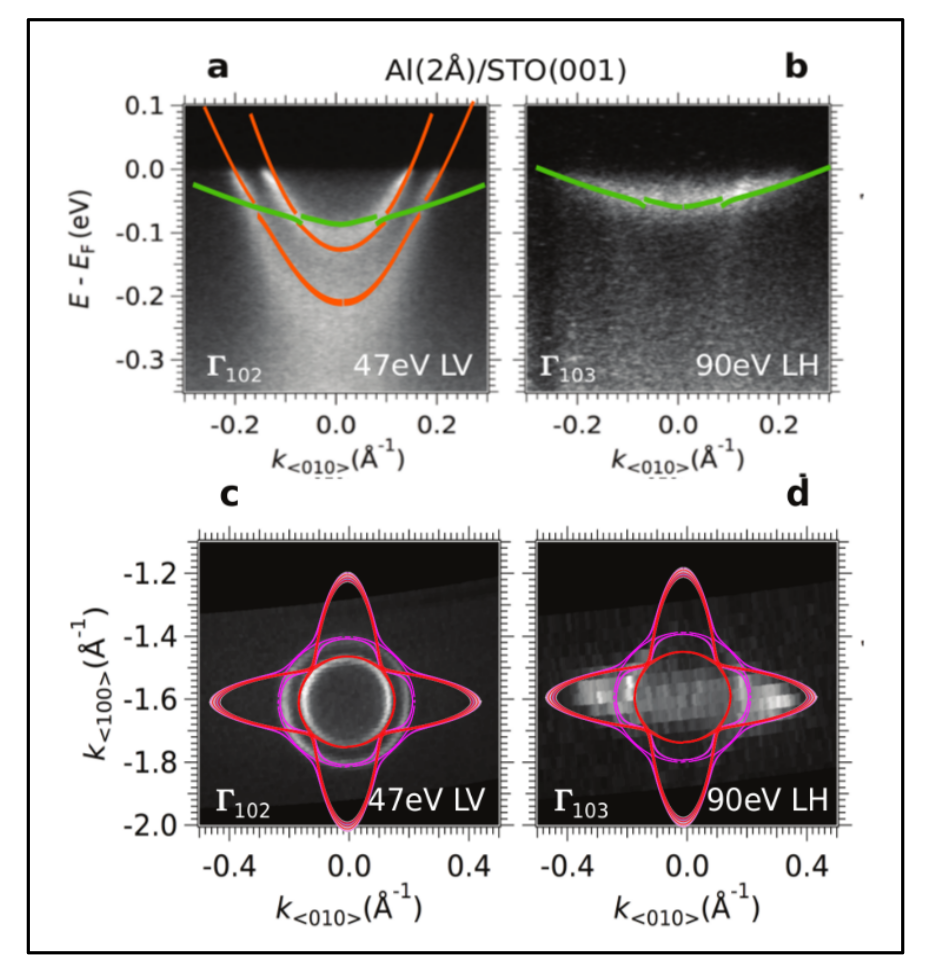}
\caption{Color online. The tight binding bands after diagonalising the Hamiltonian in Eq.(1) are overlaid with the ARPES data obtained from Ref. [\onlinecite{Roedel2016capping}] for the energy dispersion in (a) and (b) and for the fermi surface in (c) and (d). Fitting of the energy dispersion gives the parameters for the model and the accuracy of the fit is checked by overlaying the tight-binding and experimental bands at E$_f$ in (c) and (d).}
\label{fig3}
\end{figure}

\section{Nature of the eigenstates and analysis of the phase of the wavefunctions}
With the above values of the parameters, the evolution of the spectrum under the addition of each term can be seen in Fig. \ref{fig1} and Fig. \ref{fig2}. The $k$-independent bulk SOC terms and confinement produce twofold degenerate Dirac cones and another band, twofold degenerate itself, split off from the Dirac cones. 

\begin{figure}[h]
\includegraphics[width=0.5\textwidth]{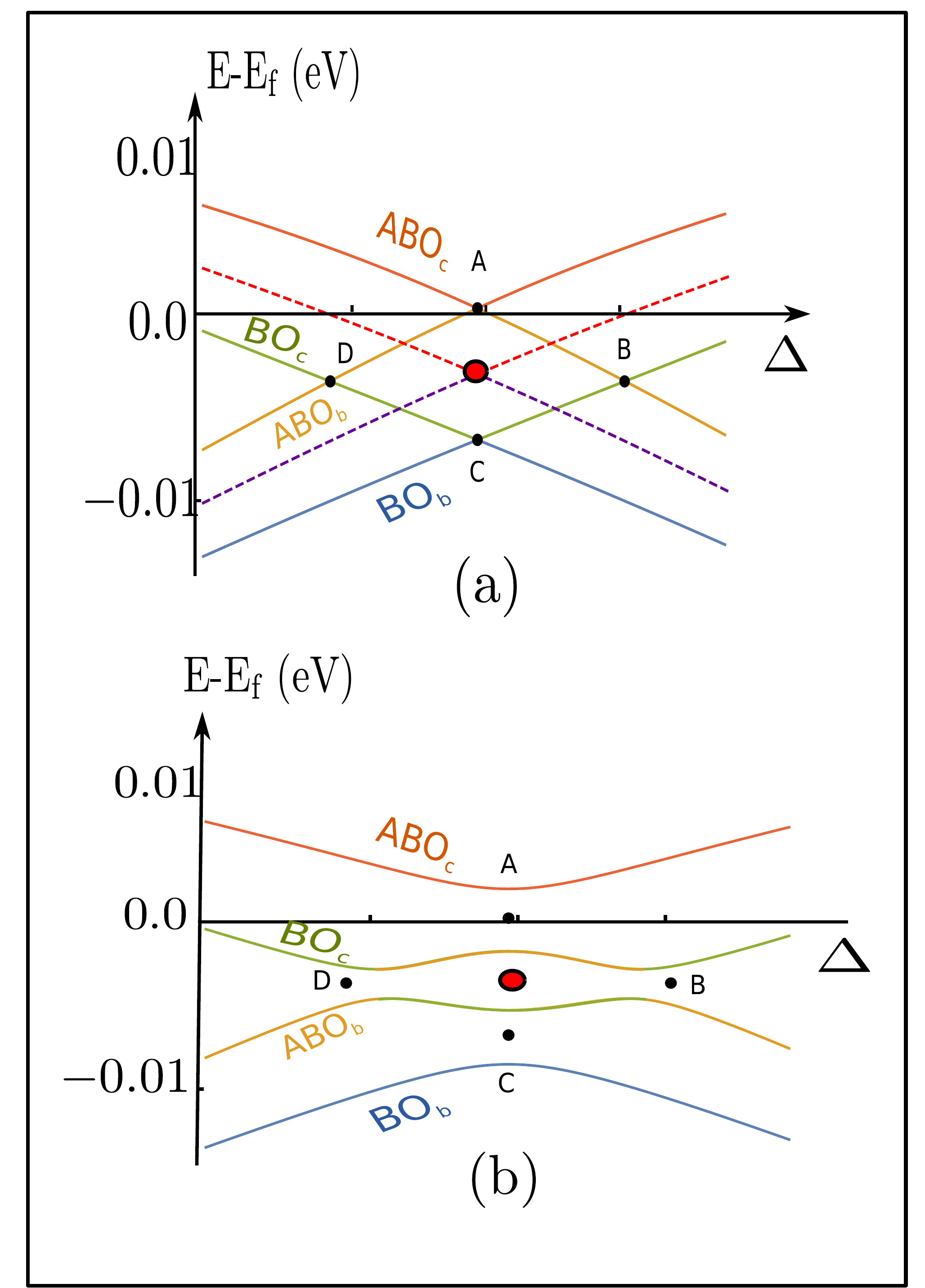}
\caption{Color online. Evolution of the doubly degenerate Dirac cone structure near $\Gamma_{\text{dp}}$ in the $\Delta$ direction, as the orbital mixing term is turned on in two steps. In (a) the branches of the Dirac cone (red dashed line  upper branch and purple dashed line, lower branch  ) split into bonding and anti-bonding lines. These cross at points A,B,C and D. In (b), \textit{g} opens up the gaps at A, B, C and D and a band inversion occurs. The parity of the bands can be seen in the colors, where green and yellow represent opposite parities.}
\label{fig4}
\end{figure}

In order to probe the effect of orbital mixing, we examine the energies and the eigenstates when $H_{\text{om}}$ is included step by step in $H_{\text{surf/2DEG}}$, beginning from $g=0$. 

a) If we set \textit{g}=0, i.e at zero orbital mixing, the spectrum in the first quadrant of the BZ, near $\Gamma_{dp}$, is plotted in Fig. 1 showing three bands, each doubly degenerate. The three corresponding eigenvalues, denoted by $E_b,E_c,E_d$, are arranged in increasing order of the energy; the first two form the tilted Dirac cones positioned about 3 meV below E$_f$, while the last one  lies 25-30 meV higher in energy above the DP and to first order, may be projected out when probing the low energy physics. We are, thus, left with twofold degenerate Dirac bands near $\Gamma_\text{dp}$. These cones can be further divided into an upper branch and a lower branch, Fig. \ref{fig4}. 
The Hamiltonian is then reduced to a block diagonal form with the upper block consisting of eigenvectors corresponding to energy $E_b$, i.e. $\Ket{B}$ and its conjugate $\Ket{B^{*}}$ and a lower block corresponding to the energy $E_c$ with eigenvectors $\Ket{C}$ and its conjugate $\Ket{C^{*}}$ . These kets are in fact, the eigen states of the Hamiltonian after addition of SOC and confinement in Eq. (\ref{ham}). After the addition of orbital mixing, we will couple these two blocks as we shall see later.
$\Ket{B}$ and $\Ket{C}$ can be represented as a linear combination of all three $d$ orbitals and can be formally noted as
\begin{eqnarray}
 \ket{B}&=a_1\ket{xy;\downarrow}+b_1\ket{yz;\uparrow}-ic_1\ket{xz;\uparrow}\\
 \ket{C}&=a_2\ket{xy;\downarrow}+b_2\ket{yz;\uparrow}-ic_2\ket{xz;\uparrow},
\end{eqnarray}
where $a_1$, $b_1$, $c_1$, $a_2$, $b_2$ and  $c_2$ are real functions of $k$ and have branch cuts.
It is to be noted that the branch cuts do not occur at the same value of the tilt for the upper and lower block.

b) The orbital mixing \textit{g} firstly contributes to a term which splits each branch of the doubly degenerate cones, upper as well as lower, into bonding and anti-bonding parts. The splitting is anisotropic and direction dependent. 

The Hamiltonian in the basis of  $\Ket{B}$,$\Ket{B^{*}}$,$\Ket{C}$ and $\Ket{C^{*}}$ reads
\begin{equation}\label{firstom}
 \begin{bmatrix}
  E_b & R_1^{*} & 0   & 0\\
  R_1 & E_b     & 0   & 0\\
  0   & 0       & E_c & R_2^{*}\\
  0   & 0       & R_2 & E_c\\
 \end{bmatrix},
\end{equation}
where $R_1=4ga_1[i\sin(k_x)b_1+\sin(k_y)c_1]$ and  $R_2=4ga_2[i\sin(k_x)b_2+\sin(k_y)c_2]$. The eigenvectors of the first block can be written in terms of the following orbitals.
\begin{eqnarray}\label{bc}
\Ket{B_1}&=\frac{1}{\sqrt{2}}e^{-i \frac{\psi_1}{2}+i\frac{\pi}{4}}\Ket{B} \\
\Ket{C_1}&=\frac{1}{\sqrt{2}}e^{-i \frac{\psi_2}{2}+i\frac{\pi}{4}}\Ket{C}
\end{eqnarray}
The eigenvectors are then given by
\begin{equation}\label{ev1}
\text{BO}_{\text{b}}=\Ket{B_1}+\Ket{B_1^{*}}
\end{equation}
corresponding to $E_b-\abs{R_1}$, and
\begin{equation}\label{ev2}
\text{ABO}_{\text{b}}=\Ket{B_1^{*}}-\Ket{B_1}
\end{equation}
for $E_b + \abs{R_1}$.
Similarly we have bonding and anti-bonding orbitals of the same character for the subspace spanned by $\ket{C},\ket{C^{*}}$,
\begin{equation}\label{ev3}
\text{BO}_{\text{c}}=\Ket{C_1}+\Ket{C_1^{*}}
\end{equation}
for $E_c -\abs{R_2}$ and
\begin{equation}\label{ev4}
\text{ABO}_{\text{c}}=\Ket{C_1^{*}}-\Ket{C_1}
\end{equation}
for $E_c+\abs{R_2}$. 

The tilted Dirac cones have a zero tilt in the $\Delta$ direction as can be seen in Fig. \ref{fig4}. The spectrum in the other directions with a non zero tilt will simply result in a rotation and shift about the DP.
Focusing on the  $\Delta$ direction for the sake of simplicity, we observe that the split branches of the cones cross at four points ABCD where we have a clear precursor of band inversion as the bonding state of the upper branch and the anti-bonding state of the lower branch intersect.  This inversion is different in nature from the orbital order reversal caused by confinement.

\begin{figure}[h]
\includegraphics[width=0.48\textwidth]{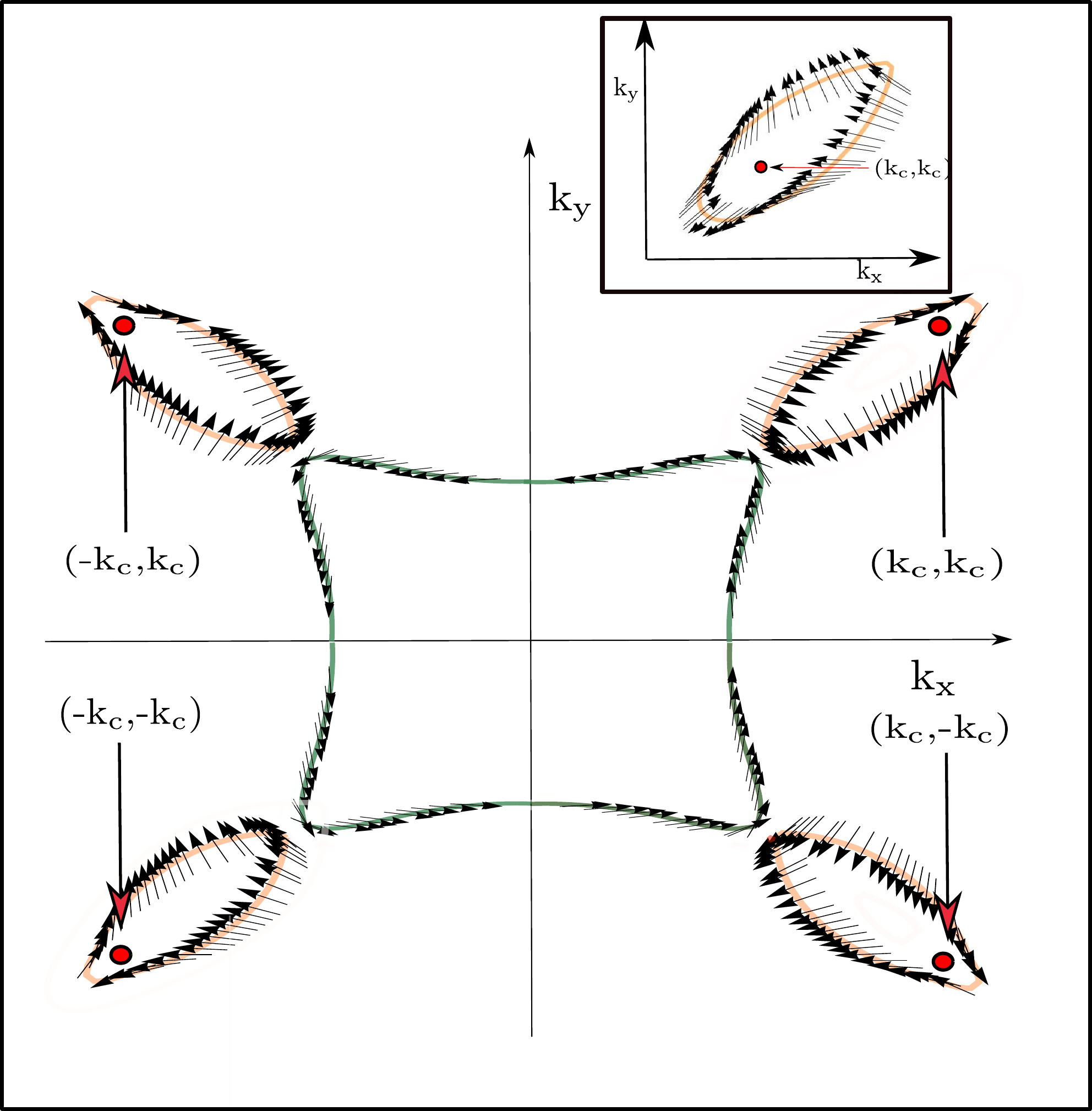}
\caption{Color online. A constant energy contour is shown for a value close to and above that of point A, after step (a) ($g=0$), see Fig. \ref{fig1}(a). One sheet is centered at $\Gamma$ and the phase $\psi$ (black arrows) changes by $+2\pi$ around the contour. Four oval shaped sheets enclose the $\Gamma_{\text{dp}}$ points (red dots). After step (b)($g\neq 0$), the phase introduced, $\psi$, winds by $-\pi$ around each contour and shows jumps of $\pi$ across the four symmetry related points. Inset: an energy cut below the point C (below $\Gamma_{\text{dp}}$) is shown.}
\label{fig5}
\end{figure}

 The orbital mixing not only splits the cones but also introduces a quantized phase difference, $\psi$, in the expressions of the eigenvectors, which can be used to characterise the Berry phase of the bands. The phases  $\psi_{1}\text{ and }\psi_2$, that enter the Eq. (\ref{ev1}-\ref{ev4}) of the eigenvectors are given by $\tan(\psi_1)=c_1\sin(k_y)/b_1\sin(k_x)$ and $\tan(\psi_2)=c_2\sin(k_y)/b_2\sin(k_x)$. The spin texture for a particular phase can then be modelled by a two component vector, as will be seen in Fig. \ref{fig5}. It is also seen that $\psi_1=-\psi_2$. The parity of the phase is opposite at points B and D as seen in Fig.\ref{fig4}, making evident the band inversion that we get even at this first step.

c) The second step of the addition of \textit{g} gives a term which gaps the spectrum at the crossing points of the anti-bonding and bonding branches, i.e. at all four points ABCD (Fig. \ref{fig2}). The phase presented above is now renormalised to $\psi^{'}$ at the points ABCD, is no longer quantized and will show jumps wherever the band gap is reached. The Hamiltonian takes the following form in the basis of the eigenvectors explicited above-

\begin{equation}\label{orbmix2}
\begin{bmatrix}
 E_b-\abs{R_1} & v\sin k_y  &w1\sin k_x &0\\
 v\sin k_y   & E_c+\abs{R_2} & 0& -w1\sin k_x\\
 w1\sin k_x &0& E_c-\abs{R_2} & v\sin k_y\\
 0 & -w1\sin k_x& v\sin k_y &E_b+\abs{R_1}
\end{bmatrix}
\end{equation}

where $w1=-2g \beta_{ab}$, $v=-2g\beta_{ac}$, $a_2b_1+b_2a_1=\beta_{ab}$ and $a_1c_2+c_1a_2=\beta_{ac}$. With the matrix elements given in Hamiltonian (\ref{orbmix2}) and that the unperturbed energies of the bonding and the anti-bonding orbitals, the perturbed energies can easily be calculated. 

The phase winding after step b) can be seen on a constant energy contour obtained right after step a) for one value slightly larger (smaller) than that at A (at C) in Fig. \ref{fig5} where the four DP, one in each quadrant, are shown for clarity. Two sets of contours are present- one centered around $\Gamma$ and four, oval shaped, lines surrounding the DP. $\psi$ is calculated around one DP in the first quadrant. For the contour above A, a branch cut extending from $\Gamma_{\text{dp}}$  to M is observed. Below  point C a branch cut extending from $\Gamma_{\text{dp}}$ to $\Gamma$ is obtained. Even though the branch cut is not along the same direction for bands with energies above and below  $\Gamma_{\text{dp}}$, the integral of the Berry phase along all the bands at a fixed value of $k$ still remains zero due to the contribution from the rest of the bands. The cones and the singularity in $\psi$ both evolve with C$_4$ rotation. Two cones which are time reversed partners have a shift of this singularity by an 
additional $\pi$ between them. 
These branch cuts are at the origin of a rather unusual behavior of the Berry phase the quantization of which is normally imposed by its single-valuedness on a closed orbit around the Dirac points. Due to the branch cut and the associated $\pi$-jump, one therefore obtains integrated values of the Berry connection that correspond to non-integer ``winding numbers'' (here half-valued). 

Below the DP, the energy contours have phases which wind in the reverse direction to that above $\Gamma_{\text{dp}}$, as can be expected. In between the points A and D, the phase also winds as expected along the energy cuts made in this region. 

According to this two-tier sequence of events, the degeneracy for the cones is lifted by $H_{\text{om}}$ and at each step one only considers the mixing of two bands at a time. While this scheme becomes more accurate as the energy difference between the upper and the lower branches increases (i.e. far from $\Gamma_{\text{dp}}$), one should, a priori,  consider all four degenerate bands instead of one pair at a time, in the vicinity of $\Gamma_{\text{dp}}$ (ABCD region). However, if we project the approximate eigenvectors obtained after the two step procedure onto the exact eigenvectors obtained by exact diagonalization of the Hamiltonian, we find an order one overlap which justifies the approximation.
It is worthy to note that the parity of the anti-bonding orbitals is always odd while that of the bonding orbitals is always even. Thus, at the crossing points B and D where the band inversion takes place, one switches from a bonding to an anti-bonding orbital at D and vice  versa at B, thus leading to an inversion in the parity of the eigenstates. With the form of the energies, which resemble gapped and inverted bands and the eigen-vectors which have an inversion of parity at the gapped points, we expect these topological properties to give rise to 1D edge states.

Next, we show that there is a convenient way to follow a variational procedure to model the Hamiltonian allowing us to describe the topological properties of the system in more detail.

\section{Edge states}
We now use a variational approach to find a model Hamiltonian, closely capturing  the properties of Eq. (\ref{ham}). The Hamiltonian, which is simply a mapping of the original problem unto a simpler and a better-understood model is a parabolic model, with breaking of the particle-hole symmetry, giving inverted bands with gap openings. Our choice of the parabolic simplification of our tight-binding bands is justified because we investigate the spectrum close to the DP, which is our new effecive "$\Gamma$" point and in its vicinity, bands can be assumed to be parabolic. The model Hamiltonian reads 

\begin{equation}\label{bhz_ham}
h(k)= \begin{bmatrix}
 \epsilon(k)+ M(k)&i\alpha k_x-\gamma k_y\\
  -i\alpha k_x-\gamma k_y& \epsilon(k)- M(k)
 \end{bmatrix}
\end{equation} 
where $\epsilon(k)=C-D_{x}k_x^2-D_{y}k_y^2\text{ and } M(k)=M -B_{x}k_x^2-B_{y}k_y^2$. The tilt of the Dirac cones, given by $(\epsilon_{d_{xy}} +\epsilon_{d_{xy}} +\epsilon_{d_{xy}})/3 $, can be expanded in the vicinity of $\Gamma_{\text{dp}}$ to give $d_0(k)\mathbf{1}=(d_{x}k_x^2+d_{y}k_y^2)\mathbf{1}$. It has been absorbed in the particle-hole-asymmetry term $\epsilon(k)$. $\alpha$ and $\gamma$ represent the anisotropy of the Rashba interaction and are themselves functions of $k$,  which incorporate the branch cuts of the eigenvectors and the associated singularity of the Berry phase. The Hamiltonian $h(k)$ in Eq. (\ref{bhz_ham}) now looks like the two band BHZ model \cite{bernevig} giving similar band inversion and gap opening. But there are also differences (i) the band gap is very small $\equiv 2$meV (ii) the Rashba coupling terms which are responsible for opening up the gap at the band crossings are anisotropic (iii) the spectrum does not have a rotational symmetry around the DP due to the tilt $d_0(k)$.

This model with band inversion at the 2D surface can then be probed for 1D edge states.
Following the method detailed in Ref. [\onlinecite{zhou2008finite}], we break translational symmetry in the $\hat{y}$ direction, while retaining it along $\hat{x}$. This leads to the spectrum (see appendix)

\begin{equation}\label{spectrum}
E=-B_{x}k_x^2+M-B_{y}\lambda_{+}\lambda_{-}+\frac{\alpha}{\gamma}B_{y}k_x(\lambda_{+}+\lambda_{-}),
\end{equation}
where the $\lambda_{\pm}$ are the roots of the secular equation $H\Ket{\psi}=E\Ket{\psi}$. In the calculation of the edge-state spectrum, we have omitted the particle-hole asymmetry, i.e. we have used $D_x=D_y=0$. Indeed, the edge states are a consequence of the underlying topology of Hamiltonian (\ref{bhz_ham}) which is not affected by terms proportional to the unit matrix. The spectrum is linear in $k_x$  and closes the energy gap at $k_x=0$.

We note that the edge states that we have described here connect bands in the vicinity of a given $\Gamma_{\text{dp}}$ point. Time reversal symmetry implies that one also expects edge states connecting bands such that one of these is located near one $\Gamma_{\text{dp}}$ point and the other is near the time-reversed $\Gamma_{\text{dp}}$ partner in the BZ (e.g. points D and D' in Fig. \ref{fig6} of the appendix). A mathematical treatment of this type of edge state is beyond the scope of the present paper. The experimental detection of these edge states will be discussed below.

\section{Discussion and Perspectives}
Our calculation suggests that the Dirac gaps of the low energy spectrum of the 2DEG are located 3 meV below E$_f$; since the resolution that can be achieved in ARPES is less than this, $\Gamma_\text{dp}$ could actually be sitting at E$_f$, but even if the equality were not strictly met, gating the sample with a few volts in a side gate geometry or with a few tens of volts in a back gate geometry will suffice to bring $\Gamma_\text{dp}$ in coincidence with the Fermi energy. However, as we pointed out, in the case of the 2DEG at the (001) surface of STO, the states that we are considering are metallic. The topological edge states that we are predicting at the 1D boundary of the sample might not be easy to detect experimentally as they could be masked by the conducting sheet. One way around this difficulty is to couple the 2DEG to conventional s-wave superconductors. This stategy has been successfully implemented in the context of HgTe/CdTe heterostructures which are known to possess 1D topological edge states. 
Experiments carried out on HgTe/CdTe quantum wells measured the Josephson radiation from gapless Andreev Bound states \cite{deacon2016josephson}. In out-of-equilibrium situations, missing Shapiro steps and emission at half of the Josephson frequency are predicted \cite{Fu2009} and they were experimentally measured \cite{Molenkamp20164piJosephson}. These are consequences of p-wave superconductivity being induced in the 2D topological insulator's edge channels when it's Josephson coupled to nearby conventional s-wave superconductors. A doublet of p-wave Andreev bound states arises and has an energy dispersion which is 4$\pi$ periodic in the Josephson phase. We suggest a similar Josephson junction setup for STO where the latter is sandwiched between two superconductors. The topological regime can be observed by gating the sample and tuning the Fermi energy to that of the edge states. A similar change in the frequency
emitted should be observed. Optimizing the carrier concentration and the width of the STO junction to reduce the number of parallel channels would allow one to evidence the topological states.

The  report of magnetism at the surface of STO raises the issue of the stability of the topological state to time reversal symmetry breaking but from an experimental standpoint, the claim of magnetism is still debatable \cite{rodel2014orientational}\cite{Baumberger2016nomagnetism}\cite{Goldhaber-Gordon2011}\cite{Taniuchi2016PEEM}. If a purported ferromagnetic exchange splitting affected the $t_{2g}$  carriers, that would be detrimental to the edge states. The DFT study of [\onlinecite{altmeyer2016magnetism}] reveals a dichotomy between localized magnetic moments of $e_g$ orbitals  and spin textures for the $t_{2g}$ conduction states. Even if spin-polarized domains were present, one might nevertheless retain some of the topological character if the state of the 2DEG at the surface were spatially inhomogeneous. According to experimental reports in STO  \cite{Taniuchi2016PEEM} and in LAO/STO \cite{Bert2011squid, Levy2014magnetism,Salluzzo2015}, polarized domains coexist with patches of metallic regions. If the 
latter percolated across the entire surface of STO, the electronic bands of the 2DEG could still present a parity inversion leading to topological 1D edge states at the boundary. 

In the present report, we have shown that a theoretical modeling of the ARPES spectra for (001) oriented STO leads us to uncover the existence of four special points in the 2D Brillouin zone of the 2DEG with energies close to $E_f$. In the vicinity of these points, the parity of the electronic wavefunctions becomes inverted. This leads to the appearance of  topological 1D states at the boundary of the 2DEG and ways to detect those are suggested.

\acknowledgements
  
Authors benefited from discussions with A.F. Santandar-Syro and E. Bocquillon. M. G. gratefully acknowledges support from the Institut Universitaire de France and from the French National Research Agency (ANR) (Project LACUNES No. 274 ANR-13-BS04-0006-01). 

\section*{APPENDIX}
\begin{appendices}

\section{Details of edge states}
\begin{figure}[h]
\includegraphics[width=0.5\textwidth]{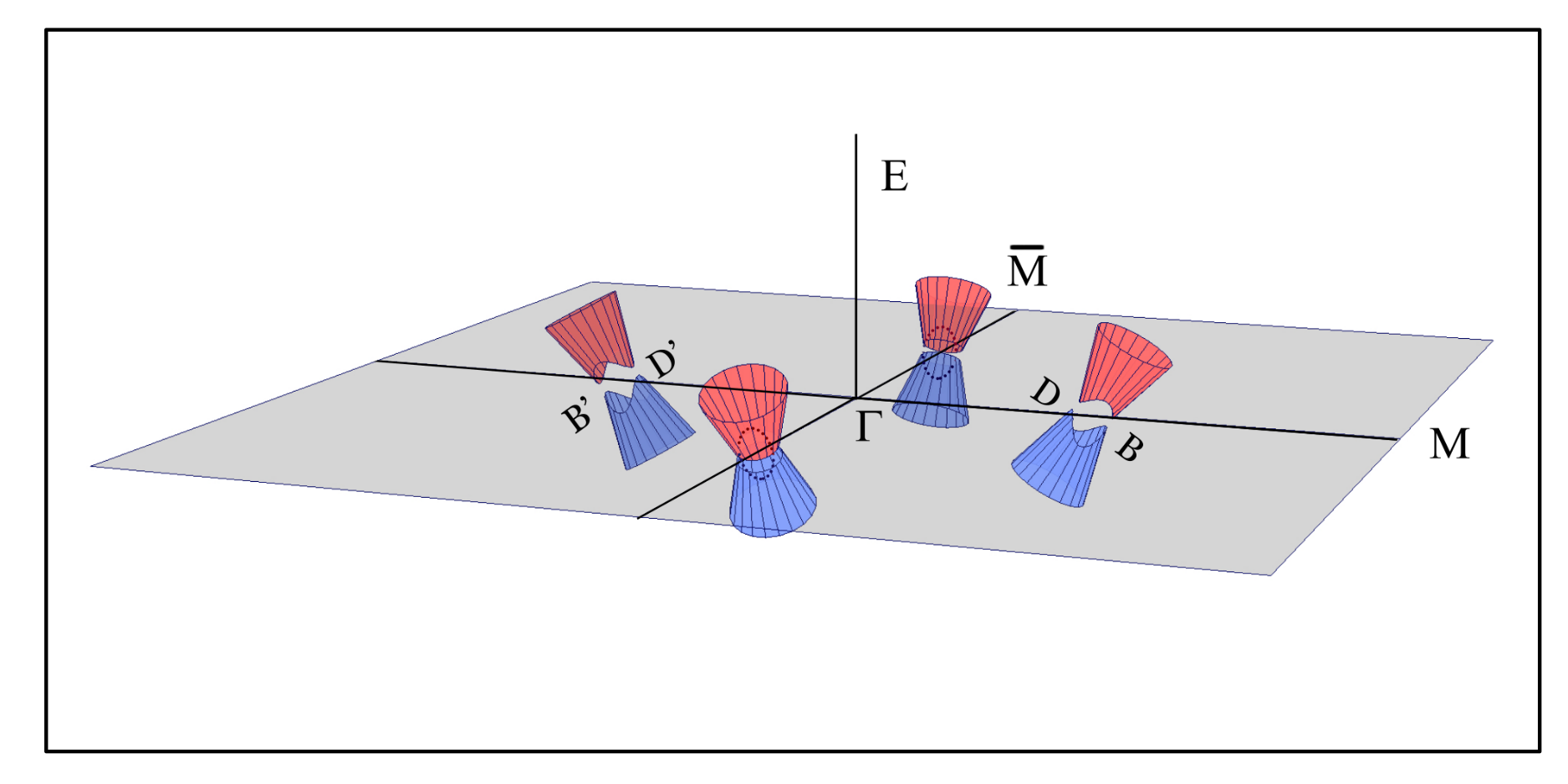}
\caption{An artist's view of the four energy bands around the Dirac points are shown in both the $\Gamma$M and the $\Gamma \bar{\text{M}}$ directions. Two kinds of edge states can be present, one between B and D and the other between D and its time reversed partner, i.e. between D$^{'}$.}
\label{fig6}
\end{figure}
The Hamiltonian \ref{bigass_ham} can be analysed either perturbatively or variationally to enable mapping onto a simpler model.
The second approach consists of using a variational approach, mapping the four band model onto an effective two band model of a similar form as used in Ref. \cite{bernevig}. However, we take explicitly into account the particle-hole asymmetry, described by the parameters $D_x$ and $D_y$ into which we have also absorbed the tilt term $d_0(k)$. The Hamiltonian, as explicited in Eq. (3) of the main text (with C=0), is
\begin{equation}\label{full_bhz}
\left[
 \begin{smallmatrix}
 M-(D_{x}+B_{x})k_x^2-(D_{y}+B_{y})k_y^2&i\alpha k_x-\gamma k_y\\
  -i\alpha k_x-\gamma k_y& -M+(-D_{x}+B_{x})k_x^2+(-D_{y}+B_{y})k_y^2\
 \end{smallmatrix}
 \right].
\end{equation}
In this section, we are interested only in the topological properties of the systems which give rise to the particular edge-state structure. Since these topological properties are not affected by the particle-hole asymmetry, which yields a term proportional to the unit matrix, we set $D_x=D_y=0$ in the following discussion of the edge states. 
The model is solved in a finite strip geometry of width $L$ with periodic (open) boundary conditions in the \textit{x}  (\textit{y}) direction. Here $k_x$ is still a good quantum number while $k_y$ is replaced by $k_y=-i\frac{\partial}{\partial y}$. The wave function is of a mixed real and $k$-space function of the form of $\phi (k_x,y)=\phi(k_x) e^{-\lambda y}$. With this form, we derive the secular equation to get four eigenvalues.
\begin{multline}\label{lamdas}
 \lambda_{\pm}^2=\frac{B_{x}}{B_{y}}k_x^2+\frac{\gamma^2}{2B_{y}^2}-\frac{M}{B_{y}} \pm \\ 
\sqrt{\left(\frac{\gamma^2}{2B_{y}^2}-\frac{M}{B_{y}}\right)^2+\frac{E^2-M^2}{B_{y}^2}+\frac{k_x^2}{B_{y}^2}\left(\frac{B_{x}}{B_{y}}\gamma^2-\alpha^2\right)}
\end{multline}
The boundary conditions at the limits impose that the solutions of this equation become zero at $y=0$ , i.e, in the middle of the slab, and go to zero as $y\rightarrow\pm\infty$.
The solutions are then of the form, for $y>0$
\begin{equation}\label{technique}
\Ket{\phi_B}=
\begin{bmatrix}
 a\\
 b
\end{bmatrix}
(e^{-\lambda_{+}y}-e^{-\lambda_{-}y}),
\end{equation}
and similarly for $y<0$.
The condition of existence of such states implies that the $\mathbbm{R}(\lambda_{+})$ and  $\mathbbm{R}(\lambda_{-})$ should be of the same sign (so as to avoid any solution which does not
decay at $\pm\infty$). The symbol $\mathbbm{R}$ denotes the real part. Under this condition, the equation $H\ket{\phi_B}=E\ket{\phi_B}$ gives us
\begin{equation}
 \frac{a}{b}=\frac{M-E-(B_{x}k_x^2-B_{y}\lambda_{+}^2)}{-i(\alpha k_x+\gamma\lambda_{+})}=\frac{M-E-(B_{x}k_x^2-B_{y}\lambda_{-}^2)}{-i(\alpha k_x+\gamma\lambda_{-})},
\end{equation}
which gives us the following spectrum
\begin{equation}
E=M-B_{x}k_x^2+\frac{\alpha}{\gamma}B_{y}k_x(\lambda_{-}+\lambda_{+})-\lambda_{+}\lambda_{-}B_{y},
\end{equation}
and we see immediately that if we are to get an edge state, we should have the energy $E=0$ at $k_x=0$, i.e., the gap is closed at $k_x=0$. For the condition  $k_x=0$, we have
\begin{equation}
\sqrt{M-E}=\text{sign}(B_{y})\sqrt{M+E},
\end{equation}
which has a solution $E=0$ for $MB_y>0$, i.e., $M$ and $B_{y}$ should have the same sign, which is the case if we have a band inversion. We also verify that in this case, we shall have 
 $\mathbbm{R}(\lambda_{+})\mathbbm{R}(\lambda_{-})>0$ and thus  $\mathbbm{R}(\lambda_{+})$ and  $\mathbbm{R}(\lambda_{-})$ have the same sign which entails $0<M/B_{y}<1/4$. Additionally, there can be two kinds of spin-polarised edge states-those present around one Dirac point, i.e. between B and D in Fig. 6 and those spanning the BZ to their time reversed partners, D and D$^{'}$.
\end{appendices}
\bibliography{bib}
\end{document}